\documentclass[prb,aps,twocolumn]{revtex4}
\usepackage{graphicx}
\begin{document}
\title{Comment on ``Strontium clusters: Many-body potential, energetics, and structural transitions'' [J. Chem. Phys. 115, 3640 (2001)]}
\author{Jonathan P. K. Doye} 
\affiliation{University Chemical Laboratory, Lensfield Road, Cambridge CB2 1EW,
United Kingdom}
\author{Florent Calvo}
\affiliation{Laboratoire de Physique Quantique, IRSAMC, 
Universit\'{e} Paul Sabatier, 118 Route de Narbonne, F31062 Toulouse Cedex, 
France}
\maketitle

\begin{figure}
\includegraphics[width=8.4cm]{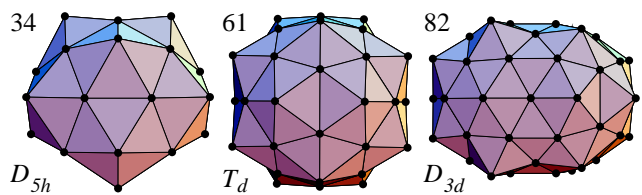}
\caption{\label{fig:structure}Possible structures for the magic number
clusters Sr$_{34}$, Sr$_{61}$ and Sr$_{82}$ along with their respective 
point groups.}
\end{figure}

In an interesting recent combined experimental and theoretical 
paper by Wang {\it et al.} on strontium clusters, the experimental
magic numbers at $N$=34 and 61 were assigned to the two structures
shown in Fig.\ \ref{fig:structure}.\cite{Wang01}
An additional strong magic number at $N$=82 remained unassigned.
As both the assigned  Sr$_{34}$ and Sr$_{61}$ structures are polytetrahedral
(that is the whole space occupied by the cluster can naturally be divided
into tetrahedra with atoms at the vertices) it seemed to us that 
a natural interpretation of the magic number at $N$=82 might be in terms of
the third structure in Fig.\ \ref{fig:structure}, which was recently
suggested as a particularly stable polytetrahedral cluster.\cite{Doye01d}

The assignment of the Sr$_{34}$ and Sr$_{61}$ structures was 
based on calculations using a model potential that Wang 
{\it et al.} parameterized.\cite{Wang01} 
However, the assigned structures were not lowest in energy, 
but calculations of the free energies of these structures (within the harmonic 
approximation) indicated that they became most stable 
at $T$=129 and 220K, respectively. 

To test our hypothesis for Sr$_{82}$ we decided to look at the 
structure and thermodynamics of this cluster using the same 
potential as in Ref.\ \onlinecite{Wang01}. As for Sr$_{34}$ and 
Sr$_{61}$ the global minimum of this cluster is not polytetrahedral,
but based on a Mackay icosahedron. However, 
using parallel tempering,
a method which has proved particularly useful for probing 
low-temperature solid-solid transitions in clusters where achieving 
ergodicity can be particularly challenging, 
we found no evidence that the 82-atom 
polytetrahedral structure depicted in Fig.\ \ref{fig:structure} 
ever became lowest in free energy.

\begin{figure}
\includegraphics[width=8.4cm]{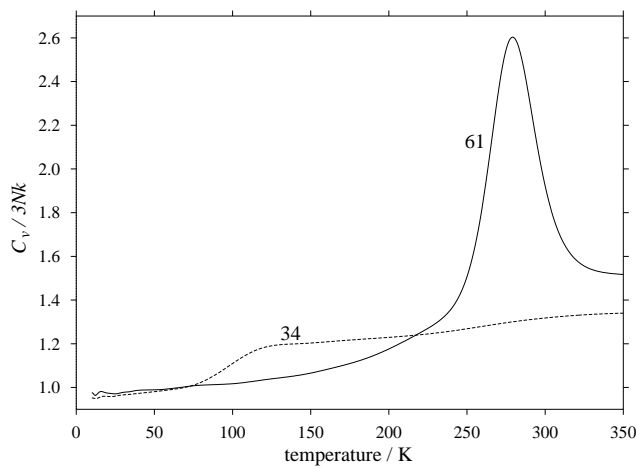}
\caption{\label{fig:thermo}Classical heat capacities for 
Sr$_{34}$ and Sr$_{61}$ calculated from parallel tempering simulations.}
\end{figure}

This failure led us to check the thermodynamics of Sr$_{34}$ and Sr$_{61}$
presented in Ref.\  \onlinecite{Wang01} using parallel tempering. 
The resulting heat capacities (Fig.\ \ref{fig:thermo}) 
are very different from Ref.\ \onlinecite{Wang01}, showing no evidence 
of large peaks in the heat capacity at the expected transition temperatures. 
Furthermore, quenches from configurations generated by the simulations
provided definitive evidence that there is no structural transition to 
the polytetrahedral structures of Fig.\ \ref{fig:structure} (or to defective
structures based upon them). 
For example, for Sr$_{61}$ at $T$=300K the $T_d$ minimum was found only 
once in 4000 quenches, compared to 59 times for the global minimum;
even above the melting temperature of this cluster, there is no sign of the 
relative stabilities of the two structures reversing.
We confirmed that these results do not reflect ergodicity problems by running simulations
starting from the $D_{5h}$ and $T_d$ structures: the clusters quickly escaped from these
structures, seldom to return. 

To identify the cause of the differences between our results and 
those of Ref.\ \onlinecite{Wang01}, we repeated Wang {\it et al.}'s 
calculations of the partition functions of the global minima and
the $D_{5h}$ and $T_d$ structures using the harmonic approximation. 
However, we found the same scenario; i.e.\ structural transitions
driven by the much greater vibrational entropy of the polytetrahedral 
structures.
That the $D_{5h}$ and $T_d$ structures are observed much 
less in the simulations than one would expect from these calculations, 
therefore points to a breakdown of the harmonic approximation as the cause of the huge 
overestimation of the partition functions of the $D_{5h}$ and $T_d$ structures.
This unusual behaviour---the harmonic approximation usually gives a correct
qualitative picture of the thermodynamics---is worthy of further investigation.
It may be that the small mean vibrational frequencies of the $D_{5h}$ and $T_d$ structures
are indicative of low barriers and basis of attractions that are small in extent.

Our results do not necessarily imply that the structures in 
Fig.\ \ref{fig:structure} are incorrect assignments for the 
experimental magic numbers. 
Although our calculations with Wang {\it et al.}'s strontium potential 
are unable to provide any evidence in support of these assignments,
this may instead reflect deficiencies in this potential and the difficulty
of reliably modelling the structure of metal clusters.

\end{document}